\documentstyle[aps]{revtex}

\begin{document}
\author{I.M. Sokolov}
\address{Theoretische Polymerphysik, Universit\"{a}t Freiburg, Hermann-Herder-Str. 3,%
\\
D-79104 Freiburg i.Br., Germany}
\title{Irreversible and reversible modes of operation of deterministic ratchets}
\date{\today}
\maketitle

\begin{abstract}
We discuss a problem of optimization of the energetic efficiency of a simple
rocked ratchet. We concentrate on a low-temperature case in which the
particle's motion in a ratchet potential is deterministic. We show that the
energetic efficiency of a ratchet working adiabatically is bounded from
above by a value depending on the form of ratchet potential. The ratchets
with strongly asymmetric potentials can achieve ideal efficiency of unity
without approaching reversibility. On the other hand we show that for any
form of the ratchet potential a set of time-protocols of the outer force
exist under which the operation is reversible and the ideal value of
efficiency $\eta =1$ is also achieved. The mode of operation of the ratchet
is still quasistatic but not adiabatic. The high values of efficiency can be
preserved even under elevated temperatures.\bigskip

{\bf PACS No: }05.70.Ln; 05.40.-a; 87.10.+e
\end{abstract}

\section{Introduction}

The general interest to the thermodynamics of living things has motivated
huge interest to the investigation of simple (probably oversimplified in
comparison with biological systems, but still very nontrivial) models of
thermodynamical systems, which are very nonlinear and are driven very far
from equilibrium. The stochastic or deterministic ratchet models present an
extremely popular class of physical models under investigation \cite
{Magnasco1,Magnasco2,Bierast,Hang1,Sekimoto,SB2,Luczka,SB1,Takagi1,Qian,A+M+P,Takagi2,Parmeggiani,Mazonka,Sokolov3,Mag+Stol,DerAst,PUK}%
. The energetic efficiency is one of the simplest (and deepest)
thermodynamical characteristics of such systems and is now under extensive
investigation \cite
{Bierast,Sekimoto,SB2,Luczka,SB1,Takagi1,Qian,A+M+P,Takagi2,Parmeggiani,Mazonka,Sokolov3,Mag+Stol,DerAst}%
. The simplest model, a rocked ratchet, corresponds to a particle moving in
a spatially asymmetric potential under the influence of the external field,
either periodic or stochastic. Under such conditions, the system generates
directed current which can flow against an additional constant external
potential difference, thus producing useful work by e.g. charging a battery.
In what follows we discuss a question of optimizing an efficiency of a
ratchet and discuss the low-temperature (deterministic) limit \cite
{Lindenb,Larralde}, as the simplest one for the optimization problem. We
show, that the efficiency of a ratchet working as a rectifier is limited by
some finite value $\eta =\eta _{\max }$, depending on the exact form of the
ratchet potential. Taking ratchet potentials which are rather flat on one
side and extremely steep on another side of a saw-tooth leads to $\eta
_{\max }\rightarrow 1$. On the other hand, under judicious choice of the
protocol of the external force eventually any cold or macroscopic ratchet
can achieve the ideal value of $\eta =1$. We show that such protocol is not
unique. Moreover, we discuss how the large values of efficiencies can be
preserved even in the systems under finite temperature.

\section{A ratchet as a rectifier}

Describing ratchet systems in a time-dependent external field one typically
starts from a Langevin equation \cite{Magnasco1},

\begin{equation}
\dot{x}=\mu \left[ F(x)+f(t)\right] +\xi (t),  \label{Lang}
\end{equation}
where $\mu $ is the mobility of the particles, and $\xi (t)$ is a $\delta $%
-correlated Gaussian Langevin force with zero mean and with $\overline{\xi
^{2}(t)}=2\theta \mu $, where $\theta =kT$ is the energetic temperature.
Here $F(x)$ is a force corresponding to the ratchet potential and $%
f(t)=f_{0}+f_{1}(t)$ is a sum of an external see-saw force $f_{1}(t)$ with
zero mean and of a constant force $f_{0}$ against which the useful work is
done by pumping the particles uphill. The evolution of the particles'
distribution $p(x,t)$ is then described by a Fokker-Planck equation,

\begin{equation}
\frac{\partial p(x,t)}{\partial t}=\frac{\partial }{\partial x}\left( D\frac{%
\partial p(x,t)}{\partial x}+\mu p(x,t)\frac{\partial }{\partial x}%
U(x,t)\right) ,  \label{FoPla}
\end{equation}
where $U(x)=V(x)-f(t)x$ is the overall potential and $D=\theta \mu $ is the
diffusion coefficient. The simplest case of a ratchet device is delivered by
a periodic piecewise-linear potential 
\begin{equation}
V(x)=\left\{ 
\begin{array}{ll}
Vx/a & \text{for }0<x\leq a \\ 
V(L-x)/(L-a) & \text{for }a<x\leq L
\end{array}
.\right.   \label{rat}
\end{equation}
Under a very slowly changing field, the system can simply be described as a
rectifier (nonlinear element) whose ''Volt-Ampere'' characteristics can
easily be calculated in an adiabatic approximation, as it was done e.g. in
Ref. \cite{Magnasco1}. The corresponding expression is given in the Appendix 
\ref{app:1}. In the limiting case of deterministic operation (corresponding
to $\theta \rightarrow 0$ or to the situation, when keeping the form of the
ratchet constant, one increases its size in a way that both $V/a$ and $%
V/(L-a)$ stay constant) the current through the ratchet as a function of the
outer field $f$ reads: 
\begin{equation}
I=\left\{ 
\begin{array}{ll}
0 & \text{for }-V/(L-a)<f<V/a \\ 
\mu \left( \frac{L-a}{f+V/(L-a)}+\frac{a}{f-V/a}\right) ^{-1} & \text{%
otherwise}
\end{array}
.\right.   \label{Iclass}
\end{equation}
Note that the deterministic current, Eq.(\ref{Iclass}) is simply given
through $I=1/\tau ,$ where 
\begin{equation}
\tau =\mu ^{-1}\left[ \frac{L-a}{f+V/(L-a)}+\frac{a}{f-V/a}\right] 
\label{time}
\end{equation}
is the time necessary for a particle to traverse the period of a ratchet.
For $-V/(L-a)<f<V/a$ the particle is trapped, and the deterministic current
vanishes: we refer to this interval as to the mobility gap. Note that the
adiabatic approximation holds if the characteristic times of varying of the
external field are much larger than $\tau $.

The general approach to energetic efficiency of a system rectifying
adiabatic field is discussed in some detail in Ref.\cite{CondMat}. In our
case it is enough to mention that the useful work, i.e. the work done
against the constant force $f_{0}$ per unit time is equal to $A_{u}=-$ $%
\overline{I(f_{0}+f_{1})f_{0}}$, and that the work of the outer force $f_{1}$
equals to $\overline{I(f_{0}+f_{1})f_{1}}$, so that the efficiency of the
ratchet is given by 
\begin{equation}
\eta =-\frac{\overline{I(f_{0}+f_{1})f_{0}}}{\overline{I(f_{0}+f_{1})f_{1}}},
\label{KPD}
\end{equation}
where the mean values are taken over the probability distribution $p(f_{1})$
of $f_{1}$ (for periodic or stochastic forces) or over a period (for
periodic ones). The energetic efficiency of a ratchet is not an intrinsic
property of the device, but depends on the protocol of the external force.
Thus, judicious choice of such protocols can improve the efficiencies
strongly.

Finding the exact upper bound of such integral expression as Eq.(\ref{KPD})
under arbitrary changes of $f_{0}$ and of the distribution of $f_{1}$ is not
an easy task even for relatively simple $I(f)$-functions. Under
deterministic operation (due to the existence of the mobility gap around $f=0
$) this optimization can follow by comparison of different situations. To do
this we note that the current, Eq.(\ref{Iclass}) is a monotonously growing
function of $f$ everywhere except of the gap $-V/(L-a)<f<V/a$. Let us
consider the case of an external force $f(t)=f_{0}+f_{1}(t)$, with a mean
value $f_{0}$ and a piecewise-constant force $f_{1}(t)$ with zero mean
taking only two values $f_{+}$ and $-f_{-}$. The value of $f(t)$ than
switches between the values $f_{\min }=f_{0}-f_{-}$ and $f_{\max
}=f_{0}+f_{+}$. Let us discuss different choices of $f_{\min },$ $f_{0}$ and 
$f_{\max }$ with respect to the gap boundaries. If both extreme values lie
within the mobility gap, no current flows and no useful work is produced.
Thus at least one of the values, $f_{\max }$ or $f_{\min }$, must lie
outside of the gap. The positive work is produced if the current $I$ has a
sign opposite to one of $f_{0}$. This can not be the case if both $f_{\min }$
and $f_{\max }$ lie on one side of the mobility gap. Thus, if e.g. $f_{\max
}>0$ lies outside of the gap, $f_{\min }$ lies either within the gap or on
its other side. In this case, if the current flows through the system, it
always flows in the direction of $f_{1}$, and thus $I(f_{0}+f_{1})f_{1}$ is
always positive. By comparing different situations it is easy to convince
oneself that under $a>L/2$ the highest efficiencies will be achieved when $%
f_{0}<0$ and $f_{\min }$ still lies within the gap (taking $f_{\min }$
outside the gap will reduce useful work and increase losses) and the value
of $f_{\max }>0$ lies outside of the gap. Having $f_{0}$ fixed, we see that
the efficiency monotonously decreases when allowing for larger $f_{1}$%
-values, since the function $I(f_{0}+f_{1})f_{1}$ always grows faster with $%
f_{1}$ than $\left| I(f_{0}+f_{1})f_{0}\right| $ does. Thus the optimization
takes place for the situation when $f_{\max }$ approaches the upper bound of
the gap from above. Similar considerations can be applied to a general case
of an external force $f(t)$ which is bounded by the values $f_{\min }$ and $%
f_{\max }$.

For the force $f_{1}$ which is symmetrically distributed around zero, i.e.
for the case $f_{+}=f_{-}$, the maximal efficiency will be achieved in the
case when $f_{0}\,$is taken near the gap's midpoint (shifted to the right by
an infinitesimally small amount) and the amplitude of $f_{1}$ is equal to
one-half of the gap's width: \smallskip 

\begin{equation}
f_{0,\max }=\frac{V(L-2a)}{2a(L-a)},\text{ }f_{1,\max }=\frac{VL}{2a(L-a)},
\label{Opt}
\end{equation}
which leads to the maximal value of the efficiency 
\begin{equation}
\eta _{{\rm sym}}=\left| 1-2a/L\right| .  \label{max}
\end{equation}
Note that the values of $f_{0,\max }$ and $f_{1,\max }$ correspond to a
stagnation situation when the mean current $\overline{I}$ just vanishes.
Thus, under any symmetric outer force the efficiency of the ratchet is
bounded by a value smaller than unity, except for the extreme cases $a=L$ or 
$a=0$. In this case the appliance shows the piecewise-linear volt-ampere
characteristics of an ideal diode. For example, for $a=L$ one gets: 
\begin{equation}
I(f)=\left\{ 
\begin{array}{ll}
0 & \text{for }f<V/L \\ 
\mu (f-V/L) & \text{otherwise}
\end{array}
.\right.   \label{extreme}
\end{equation}
The case $a=0$ corresponds to a symmetric situation, $f\rightarrow -f$.

The use of asymmetric external force leads to the efficiencies which are
somewhat higher, but still not arbitrarily close to unity. Taking both $%
f_{\min }$ and $f_{0}$ to lie within the gap, near to its lower boundary $%
f_{\min }\approx f_{0}\approx -V/(L-a)$, and $f_{\max }$ being slightly
higher then the upper boundary of the gap, we get that the value of
efficiency tends to 
\begin{equation}
\eta \simeq \frac{-f_{0}}{f_{\max }-f_{0}}\simeq \frac{V/(L-a)}{V/a+V/(L-a)}%
=a/L.
\end{equation}
In this case the force $f_{1}$ is strongly asymmetric: it takes almost all
the time the small negative value of $-f_{-}$, and switches from time to
time to a positive value of $f_{+}$ which equals to the gap's width. Note
that for the case $a<L/2$ one shifts the value of $f_{0}$ and $f_{\max }$ to
the right boundary of the gap and take $f_{\min }$ slightly smaller than its
left boundary, so that in general $\eta _{\max }=\max \left\{
a/L,(L-a)/L\right\} =1/2+\left| 1/2-a/L\right| $. Introducing the asymmetry
parameter $\epsilon $ through $a=L(1+\epsilon )/2$, one gets that the
maximal efficiency for a symmetric outer force is $\eta _{{\rm sym}}=\left|
\epsilon \right| $, while the absolute maximum of the efficiency (for
strongly asymmetric force) reads $\eta _{\max }=1/2+\left| \epsilon \right|
/2$.

The fact that the adiabatic efficiencies do not achieve unity is easy to
explain within the picture of deterministic motion. Let us consider
meandering dichotomic external force switching between the values -$f_{-}$
and $f_{+}$. The particle in such a system is moving to the right when $%
f=f_{+}$ and gets stuck otherwise (it never steps backwards, cf. Ref.\cite
{PUK}). The highest efficiency is achieved under stalling condition, so that
for $f=f_{+}$ one side of the saw-tooth (say, the left one of the length $a$%
) gets practically horizontal. The velocity of a particle is infinitesimally
small when passing this side; the overall passing time diverges and the
current vanishes. On the other hand, the motion of the particle along the
other side of the ratchet, which is rather steep, follows with a finite
velocity and is thus connected with finite losses, showing that the overall
process is irreversible. The heat produced when sliding down the steep side
is equal $Q=V[(L-a)/a+1]$ per particle and does not vanish when $%
a\rightarrow L$. On the other hand, the maximal useful work per particle
done during the time the particle traverses the period of the potential is
given (for the case of the forces, Eq.(\ref{Opt})) by $A=V[L(2a-L)]/[2a(L-a)]
$ and grows when $a\rightarrow L$ due to the possibility to increase the
amplitude of the external force. The value of $\eta =A/(A+Q)$ then
approaches unity not because the losses vanish but because the work $A$
grows, i.e. {\it without reaching reversibility}. On the other hand, as we
proceed to show, the reversible operation is also possible for each ratchet.

\section{The reversible mode of operation}

The analysis of the previous Section gives an idea how to reduce the losses:
as soon as the particle crosses the apex of the potential, the see-saw force
must change its sign, so that the velocity of the particle's falling down
stays infinitesimally small. The losses will be minimized while the useful
work stays finite. This will correspond to the reversible mode of operation.

Let us discuss a periodic outer field (period $T$) and concentrate on the
situation when the particle's displacement per period of the outer force
corresponds to the period of the ratchet force $L$. Note that this situation
never leads to adiabatic behavior. The useful work per period of the outer
force is constant and is given by $A=f_{0}L$. Thus, for a given ratchet
force $F(x)$ we are to look for a protocol (time-dependence) of $f_{1}(t)$
which minimizes the heat $Q=\int_{0}^{T}\mu ^{-1}v^{2}(t)dt$, where the
particle's velocity $v(t)=\dot{x}(t)$ is a periodic solution of a nonlinear
differential equation

\begin{equation}
\dot{x}(t)=\mu \left( F(x_{0}+x(t))+f_{0}+f(t)\right)   \label{Main}
\end{equation}
(being Eq.(\ref{Lang}) without the noise term) for a given initial condition
(say $x_{0}=0$). The efficiency of the ratchet $\eta =A/(A+Q)$ will then
approach unity when the heat $Q$ vanishes. Since $Q=\mu ^{-1}\overline{%
v^{2}(t)}T$, the heat can vanish only for those protocols for which $%
\overline{v^{2}(t)}$ tends to zero, i.e. only under quasistatic conditions.

For example, confining ourselves to piecewise-linear ratchet potentials and
piecewise-constant force $f_{1}(t)$ one finds that the duration of
corresponding sub-periods $t_{1}$ and $t_{2}$ ($t_{1}+t_{2}=T$) must be
equal to the time necessary for a particle to pass the distance $%
a=L(1+\epsilon )/2$ and $L-a=L(1-\epsilon )/2$, respectively. Let us
introduce a temporal asymmetry parameter $\delta $ of the external force so
that $t_{1}=T(1+\delta )/2$. Supposing that the mean value of $f_{1}(t)$ is
zero, we get $f_{-}=f_{+}(1+\delta )/(1-\delta )$. According to Eq.(\ref
{Main}) during the first sub-period $t_{1}$ the particle moves with the
velocity $v_{1}=\mu \left( -2V/L(1+\epsilon )+f_{0}+f_{+}\right) $, while
during the second sub-period $t_{2}$ it moves with the velocity $v_{2}=\mu
\left( 2V/L(1-\epsilon )+f_{0}-f_{-}\right) $. In the optimal mode of
operation the both velocities have to vanish, from which the values of $f_{0}
$ and $f_{1}$ follow: 
\begin{equation}
f_{0}=-\frac{2V(\epsilon -\delta )}{L(1-\epsilon ^{2})}\text{ and }f_{+}=%
\frac{2V(1-\delta )}{L(1-\epsilon ^{2})}.
\end{equation}
The positive work is produced if $f_{0}<0$. For example, a temporary
symmetric situation ($\delta =0$) would correspond to $f_{1}=\pm
2V/L(1-\epsilon ^{2})$. The mode of operation in this case is reversible:
the work per particle per period of the field stays finite when increasing
the period, while the losses vanish. In contrast with a rather robust
adiabatic situation of the previous Section, achieving high efficiency in a
reversible regime assumes fine-tuning of the temporal parameters of the
field: the high efficiencies follow as a kind of a nonlinear resonance. In
Appendix \ref{app:2} a way for finding such protocols for the potentials of
arbitrary form is discussed.

Note that the solution of Eq.(\ref{Main}) follows under the initial
condition $x(0)=0$. For different initial conditions the protocol of the
outer force has to be chosen in a different way (at least shifted in time).
On the other hand, for a wide class of protocols the particle gets
resynchronized with the field within finite time under eventually any
initial position. For example, our rectangular meandering protocol for the
piecewise-linear ratchet belongs to this class: if the particle starts at $%
t=0$ with some coordinate $x>0$, it will reach the point $x=L$ earlier in
time and will stay there until the field will be switched in the opposite
direction, thus resynchronizing its position. Another solution uses the
possibility to synchronize the external field with the instantaneous state
of the system, for example, to change the field's direction exactly at the
moment when the particle crosses the cusp of the ratchet potential. This
kind of synchronization corresponds to an engineering solution widely used
since it was first introduced by a boy called Humphrey Potter into a
Newcomen's steam engine in 1713 (see e.g. Ref. \cite{steam} as an excellent
old source). Humphrey Potter connected the cock regulating the access of
steam into the cylinder with one of the moving parts of the engine and thus
was able to increase the velocity of the operation by the factor of two.
This is exactly the kind of synchronization implied e.g. by the mode of
operation of a model-B ratchet of Ref.\cite{Parmeggiani}: in a biological
system one can imagine that a particle catalyzes the process leading to
triggering of the external field, when it approaches an active center
located near the cusp of the potential. Note that the systems which are
synchronized in this sense can also be called the systems with localized
transitions (which spatial aspect is strongly emphasized in Ref.\cite
{Parmeggiani}). Although the one-particle appliances with localized
transitions may stay reversible even under finite temperatures (just line
the appliance of Ref.\cite{Parmeggiani} does), this way of achieving high
efficiencies is not general: The Humphrey Potter's solution seems to be
ideal only in one-particle appliances, since it is impossible to synchronize
the external force (described by a single variable) with a state of a system
of several particles described in a multidimensional phase space. Thus,
either these particles must be connected in such a way that they build a
macroscopic system described by a single effective variable, or the
triggering must take place when a considerable amount of particles gather in
the vicinity of the active center.

\section{Operation under noisy conditions: size vs. temperature}

The resonant character of the reversible mode of operation makes it very
sensitive to perturbations, e.g. to the thermal noise. Thus, for achieving
ideal efficiencies the system must be taken to be cold or macroscopic. In
what follows we make some estimates, how cold or how macroscopic it has to
be.

Imagine a particle situated at $t=0$ at a lower cusp of the saw-tooth, $x=0$%
. Statistical fluctuations connected with thermal motion lead to the fact
that during the time necessary for a deterministic trajectory to pass the
apex of potential at $x=a$, in approximately one half of all realizations
the particle still not reaches the apex, so that a half of trajectories does
not contribute to a useful work at all. In the other half of realizations
the particle has already passed the apex, so that is moves with nonzero
velocity and contributes to losses. 

Let us vary the protocol slightly in a way that warrants that a large amount
of particles crosses the apex of the potential when the force is showing in
the correct direction. On the other hand, the mean square velocity of the
particles is to be kept as small as possible to reduce losses. In order to
increase the probability that the particle starting at zero crosses the apex
during the time the force is showing, say, to the right, let us increase $%
t_{1}$ by the amount $\Delta t$ necessary to guarantee crossing the top even
when the distribution of the particle's position gets broadened by
diffusion. Let us consider a ratchet whose asymmetry factor $\epsilon $ is
not extreme (not too near to zero or unity), so that both slopes are of the
same order of magnitude, $\left| F(x)\right| \simeq U/L.$ The lengths of
both sides are of the order of magnitude of $L$. During the time $%
t_{1}=a/v_{1}$ of travel along the left part of the saw-tooth with a
constant velocity $v_{1}$, the distribution of the particle's positions
broadens to the width $W\simeq \sqrt{Dt_{1}}$. Thus, waiting an additional
time $\Delta t\simeq W/v_{1}$ assures that the apex will be crossed in
almost all realizations. If we start from a well-localized situation, then $%
\Delta t\simeq \sqrt{Da/v_{1}}/v_{1}=\sqrt{Da}v_{1}^{-3/2}$. Having crossed
the apex the particles slide along the steep part of the saw-tooth with the
velocity $v_{f}\simeq \mu F\simeq \mu U/L$. The overall losses during the
first half-period of outer force are the of the order of $v_{1}^{2}t_{1}/\mu
+v_{f}^{2}\Delta t/\mu \sim v_{1}a/\mu +v_{f}^{2}\sqrt{Da}v_{1}^{-3/2}/\mu $%
. After most of the particles have passed the apex, the outer field is
switched to the opposite direction, and the particles move along the other
side of the saw-tooth with small velocity $v_{2}$. Let us take $v_{1}$ and $%
v_{2}$ to be of the same order of magnitude: $v_{1}\simeq v_{2}\simeq v$.
For the overall losses we get: $Q\simeq BvL/\mu +C\mu
U^{2}L^{-3/2}D^{1/2}v^{-3/2}$ (where $B$ and $C$ are some numerical factors
depending on the form of the ratchet), which expression is minimized under $%
v\sim \mu U^{4/5}\theta ^{1/5}L^{1/5}$. This leads by the order of magnitude
to $Q\simeq U^{4/5}\theta ^{1/5}$, while the useful work is of the order of $%
A=f_{0}L\simeq U$, so that 
\begin{equation}
1-\eta \simeq (U/\theta )^{-1/5}.  \label{hot}
\end{equation}
Thus, the way to increase the efficiency of the system (i.e. to make $Q$
small compared to $A$) is o take $U/\theta \gg 1$, i.e. to make system
macroscopic or cold. Note that if the typical value of the external field is
given, the saw-tooth height $U$ can be made large compared to $\theta $ by
introducing flat, long steps, thus leading to high efficiencies. On the
other hand, the dependence of $\eta $ on size or temperature is weak
(follows a power-law with a very small exponent of 1/5) so that the high
efficiencies can be only achieved on rather macroscopic scales. Note also
that at given size and temperature the smallest losses are achieved under
finite velocity, i.e. under the {\it finite-time} mode of operation. The
typical period of the field will scales as $t\sim L/v\sim \mu
^{-1}L^{4/5}\theta ^{-1/5}U^{-4/5}$ and diverges under $\theta \rightarrow 0$
or $L\rightarrow \infty $.

We have discussed the situation when at the beginning of the period the
particles were localized at the lower cusp of a saw-tooth, near $x=0.$ Thus,
in order to allow for periodic operation, one has to give the particles
enough time to collect into a narrow region near the bottom of the potential
after performing the cycle of operation (which corresponds to a period of
the external field). To do this one can add to the time necessary for making
the way from the top to the bottom of the potential an additional amount of
time necessary to thermalize, and to reduce the state to its initial width.
For example, one can add at the end of each period an amount of time $t_{0}$
during which the see-saw force is switched to zero. Then each new period
will start from a distribution of particles localized in a spatial domain of
width $W_{1}\sim (\theta /U)L$, centered at $x\approx 0$. The magnitude of
the maximal efficiency achieved under this procedure is still given by Eq.(%
\ref{hot}).

\section{Conclusions}

Driven ratchets belong to the class of strongly nonlinear and strongly
nonequilibrium thermodynamical systems, so that the problem of optimizing
their efficiencies is not simple. We have considered the simplest situation
of a cold or macroscopic ratchet and discussed its efficiency under
different conditions. Since any finite-velocity mode of operation leads to
inevitable losses, the highest efficiencies are achieved in quasistatic
regime. We have shown that efficiency of a ratchet as a rectifier, i.e. a
device transforming adiabatically slowly changing outer field with zero mean
into a continuous directed motion, is bounded by some value $\eta _{\max }$
depending on the geometry of the ratchet and the form of the outer force.
For ratchets consisting of extremely steep and of almost flat parts $\eta
_{\max }\rightarrow 1$: the system approaches ideal efficiency in the
irreversible mode of operation. On the other hand, we have also shown, that
for a wide class of (nonrandom) protocols of the outer force a reversible
mode of operation is possible, under which the ideal efficiency of $\eta =1$
is also achieved. This mode corresponds to a quasistatic but not adiabatic
operation. It is worth to note that such a synchronized ratchet resembles
much more a well-constructed engine, than an appliance extracting work from
a noise; in macroscopic systems high efficiencies of reversible operation
are preserved also at elevated temperatures.

\section{Acknowledgments}

The author is thankful to J. Klafter, A. Blumen and U. Handge for
enlightening discussions. Financial support by the Deutsche
Forschungsgemeinschaft through the SFB 428 and by the Fonds der Chemischen
Industrie is gratefully acknowledged.

\newpage

\appendix

\section{The Volt-Ampere characteristics of a ratchet}

\label{app:1}

In an adiabatic regime the current $I$ through the ratchet is given by a
stationary solution of the Fokker-Planck equation, Eq.(\ref{FoPla}): 
\begin{equation}
D\frac{dp(x)}{dx}+\mu p(x)\frac{d}{dx}U(x)=-I.
\end{equation}
with $D=\mu \theta $. The formal solution of this equation reads
\begin{equation}
p(x)=\exp (-U(x)/\theta )\left[ p(0)-(I/\mu \theta )\int_{0}^{x}\exp
(U(x^{\prime })/\theta )dx^{\prime }\right] .  \label{DGL1}
\end{equation}
The current $I$ and the integration constant $p(0)$ follow then from the
additional conditions representing the periodicity $p(L)=p(0)$ and the
overall normalization $\int_{0}^{L}p(x)dx=1$, leading to: 
\begin{eqnarray}
I &=&\mu \theta \left[ \frac{e^{-U(L)/\theta }}{e^{-U(L)/\theta }-1}\left(
\int_{0}^{L}dx^{\prime }e^{-U(x^{\prime })/\theta }\right) \left(
\int_{0}^{L}dx^{\prime }e^{U(x^{\prime })/\theta }\right) \right. - \\
&&-\left. \int_{0}^{L}dx^{\prime }e^{-U(x^{\prime })/\theta
}\int_{0}^{x^{\prime }}dx^{\prime \prime }e^{U(x^{\prime \prime })/\theta
}\right] ^{-1}.  \nonumber
\end{eqnarray}
The solution for the piecewise-linear potential, Eq.(\ref{rat}) is given by
the following expression:

\begin{eqnarray}
\mu I^{-1} &=&\frac{L-a}{f+V/(L-a)}+\frac{a}{f-V/a}+ \\
&&\theta \left\{ \frac{\exp \left[ -\left( V+f(L-a)\right) /\theta \right] -1%
}{\left[ f+V/(L-a)\right] ^{2}}-\frac{\exp \left[ -\left( V-fa\right)
/\theta \right] -1}{(f-V/a)^{2}}\right\} +  \nonumber \\
&&\theta \left\{ \frac{\exp \left[ -\left( V-fa\right) /\theta \right] -1}{%
f-V/a}-\frac{\exp \left[ -\left( V+f(L-a)\right) /\theta \right] -1}{%
f+V/(L-a)}\right\} ^{2}\times   \nonumber \\
&&\times \left\{ \exp \left[ -\left( V-fa\right) /\theta \right] -\exp
\left[ -\left( V+f(L-a)\right) /\theta \right] \right\} ^{-1}.  \nonumber
\end{eqnarray}
For $\theta \rightarrow 0$ one of the exponential terms is positive and
large for $f>V/a$ and for $f<-V/(L-a)$, and both of them tend to zero for $%
-V/(L-a)<f<V/a$, in which case the current vanishes. Thus, in this limiting
case Eq.(\ref{Iclass}) follows. Note that the same limiting behavior emerges
when, keeping the form of the ratchet constant, one increases its size in a
way that both $V/a$ and $V/(L-a)$ stay constant.

\section{The best protocol for a given ratchet}

\label{app:2}

Let us discuss the procedure of finding good protocols for reversible
operation for a given ratchet potentials of general form. Let us fix some
differentiable, monotonously growing function $\xi (\tau )$, $0\leq \tau
\leq 1$, so that $\xi (0)=0$ and $\xi (1)=1$. Note that under these
conditions an inverse function $\tau (\xi )$, exists, which is also a
monotonously growing, differentiable function. Let us now consider $\xi $ as
the particle's relative position within the period of potential $L$ and $%
\tau $ as the reduced time, so that $x(\tau T)=L\xi (\tau )$. Let us now
look for a protocol that realizes the corresponding $x(t)$ motion, and
increase the period of the field. From Eq.(\ref{Main}) one then gets: 
\begin{equation}
f(T\tau )=-F(L\xi (\tau ))-f_{0}+\frac{L}{\mu T}\frac{d\xi }{d\tau }
\label{Rev}
\end{equation}
Eq.(\ref{Rev}) allows finding good protocols for a given ratchet or good
ratchets for given protocol. Requiring the zero mean value of the see-saw
force one gets: 
\begin{equation}
f_{0}=-\int_{0}^{1}F(L\xi (\tau ))d\tau +\frac{L}{\mu T}=\frac{L}{\mu T}%
-\int_{0}^{1}F(L\xi )\frac{d\tau }{d\xi }d\xi
\end{equation}
Since the positive work requires $f_{0}<0$, each protocol, for which 
\begin{equation}
\int_{0}^{1}F(L\xi )\frac{d\tau }{d\xi }d\xi >0  \label{Positive}
\end{equation}
tends to be a good one when performed slowly enough. Since the losses are $%
Q=\mu ^{-1}\int_{0}^{T}v^{2}(t)dt$, and since $v(t)=(L/T)\xi ^{\prime }(t/T)$%
, one has

\begin{equation}
Q=\mu ^{-1}\frac{L^{2}}{T}\int_{0}^{1}(\xi ^{\prime }(\tau ))^{2}d\tau \text{%
.}
\end{equation}
The corresponding integral is supposed to converge: 
\begin{equation}
\kappa =\int_{0}^{1}\left( \xi ^{\prime }(\tau )\right) ^{2}d\tau <\infty .
\label{Finite}
\end{equation}
The exact behavior of this function is not very important. The efficiency of
the corresponding engine will then be 
\begin{equation}
\eta =\frac{\left| f_{0}\right| }{\left| f_{0}\right| +\kappa \mu L/T}
\end{equation}
which tends to unity when the period of the external force grows. Thus, each
reversible function $\xi (\tau )$ fulfilling Eqs.(\ref{Positive}) and (\ref
{Finite}) delivers a good protocol for a reversible operation.

For example let us take an arbitrary positive, monotonously increasing
bounded function $g(x)$. Then taking $\tau (\xi )=\int_{0}^{\xi }g(F(L\zeta
))d\zeta /\int_{0}^{1}g(F(L\zeta ))d\zeta $ guaranties that $%
\int_{0}^{1}F(L\xi )\frac{d\tau }{d\xi }d\xi >0$, since the regions where $%
F(L\xi )$ is positive get in the integral the higher weight than those where
it is negative and since $\int_{0}^{1}F(L\xi )d\xi =0$. This class of
protocols does not cover the whole set of possible protocols, since many
other protocols, like piecewise linear functions $\tau (\xi )=(\alpha +\beta
\Theta (F(L\xi ))\xi \,$(where $\alpha ,\beta >0$ and $\Theta $ is a
Heaviside step-function) will also lead to eventually ideal mode of
operation. A protocol for the piecewise-linear ratchet discussed in Sec.III
corresponds just to this class.

\end{document}